\documentclass[preprint,showpacs,preprintnumbers,amsmath,amssymb]{revtex4}

\usepackage[dvips]{graphicx}
\usepackage{dcolumn}
\usepackage{bm}
\begin{document}

\title{Unusual superexchange pathways in a Ni triangular lattice of NiGa$_2$S$_4$
with negative charge-transfer energy}

\author{K. Takubo}
\affiliation{Department of Physics \& Department of Complexity Science and Engineering, 
University of Tokyo, Chiba 277-8561, Japan}
\author{T. Mizokawa}
\affiliation{Department of Physics \& Department of Complexity Science and Engineering, 
University of Tokyo, Chiba 277-8561, Japan}
\author{J.-Y. Son}
\affiliation{Department of Physics \& Department of Complexity Science and Engineering, 
University of Tokyo, Chiba 277-8561, Japan}
\author{Y. Nambu}
\affiliation{Department of Physics, Kyoto University, Kyoto 606-8502,Japan}
\affiliation{Institute for Solid State Physics, University of Tokyo, Chiba 277-8581, Japan}
\author{S. Nakatsuji}
\affiliation{Department of Physics, Kyoto University, Kyoto 606-8502,Japan}
\affiliation{Institute for Solid State Physics, University of Tokyo, Chiba 277-8581, Japan}
\author{Y. Maeno}
\affiliation{Department of Physics, Kyoto University, Kyoto 606-8502,Japan}

\date{\today}

\begin{abstract}
We have studied the electronic structure of the Ni triangular lattice
in NiGa$_2$S$_4$ using photoemission spectroscopy and subsequent model 
calculations. The cluster-model analysis of the Ni 2$p$ core-level 
spectrum shows that the S 3$p$ to Ni 3$d$ charge-transfer energy 
is $\sim$ -1 eV and the ground state is dominated by the $d^9L$
configuration ($L$ is a S 3$p$ hole).
Cell perturbation analysis for the NiS$_2$ triangular lattice
indicates that the strong S 3$p$ hole character of the ground state
provides the enhanced superexchange interaction between 
the third nearest neighbor sites.
\end{abstract}

	
\maketitle

\begin{figure}
\begin{center} 
\includegraphics[width=7cm]{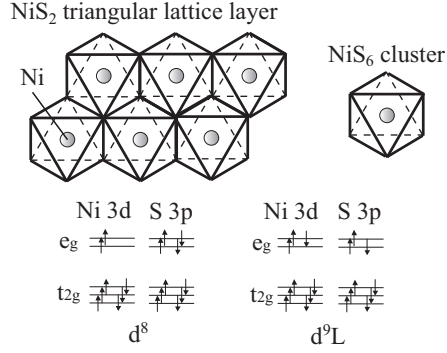}
\caption{Schematic drawing for the Ni$^{2+}$ ($S$=1) triangular lattice
layer and the electronic configurations of the NiS$_6$ cluster model
used to analyze the photoemission spectra. The NiS$_2$ layer is
constructed from the NiS$_6$ octahedra in which one Ni ion is
located at the center and six S ions at the verteces. 
In the NiS$_6$ cluster, the S 3$p$ molecular orbitals 
with $e_g$ and $t_{2g}$ symmetry hybridize with
the Ni $3d$ orbitals with $e_g$ and $t_{2g}$ symmetry.
$L$ denotes a hole in the S 3$p$ molecular orbitals.
The charge-transfer energy $\Delta$ is given by
the excitation energy from $d^8$ to $d^9L$.}
\label{structure}
\end{center}
\end{figure}

Unusual magnetic properties of geometrically frustrated 
spin systems have been attracting broad interest
of many theorists and experimentalists 
in the field of condensed matter physics
\cite{Moessner1}.
In frustrated spin systems with orbital
degeneracy, specific orbital patterns 
can lift the magnetic frustration and 
some long range magnetic orderings 
can be realized in the ground state
\cite{Pen,Lee1,Motome,Schmidt,Khomskii}. 
In frustrated spin systems without orbital degeneracy,
spin-lattice coupling tends to provide 
cooperative distortion lifting the magnetic 
frustration \cite{Lee2,Yamashita,Tchernyshyov}
or small randomness tends to induce spin
freezing at low temperature \cite{Ramirez}.
Therefore, although spin disordered states
including the resonating-valence bond state
are expected in the geometrically frustrated 
spin systems \cite{Moessner2,Canals}, 
few systems show spin disordered 
states as ground states \cite{Shimizu}.
Among the various magnetic compounds
with geometrically frustrated lattice, 
the newly-discovered NiGa$_2$S$_4$ has
the Ni$^{2+}$ ($S$=1) triangular lattice layer
without orbital degeneracy (see Fig. \ref{structure})
and is found to 
have a frozen spin-disordered ground state using neutron diffraction and 
nuclear quadrupole resonance experiments \cite{Nakatsuji,Takeya}.
The neutron result also indicates that the spin-spin
correlation between the third nearest neighbors
is much stronger than that between the first
and second nearest neighbors,
indicating that the conventional triangular lattice
model with the nearest neighbor superexchange coupling
is not enough to describe NiGa$_2$S$_4$ \cite{Fujimoto}.

In order to understand the origin of the unusual magnetic
properties of NiGa$_2$S$_4$, it is highly important
to reveal its underlying electronic structure. 
In this Letter, we report photoemission study of 
NiGa$_2$S$_4$ single crystals. The photoemission results
and subsequent model calculations indicate that the
ground state has the $d^9L$ character ($L$ is a S 3$p$ hole)
and that the strong S 3$p$ hole character of the ground state
provides the enhanced superexchange interaction between 
the third nearest neighbor sites. NiGa$_2$S$_4$ is
a unique spin-disordered system in that the negative
charge-transfer energy allows relatively long 
superexchange pathways. 

Single crystals of NiGa$_2$S$_4$ are grown by chemical vapor transport
as described in ref. \cite{Onuma}.
The x-ray photoemission spectroscopy measurements were performed 
using a JPS 9200 spectrometer equipped with a monochromatized Al $K\alpha$ 
x-ray source ($h\nu$ = 1486.6 eV). 
The total energy resolution was $\sim$ 0.6 eV.
The single crystals were cleaved {\em in situ} in order
to obtain clean surfaces. All photoemission data were collected 
at room temperature within 48 hours after cleaving.
As expected from the layered structure of NiGa$_2$S$_4$, the obtained 
surfaces were extremely clean and stable: the photoemission spectra were
consistent with the stoichiometric surface and did not show any
change during the experiment.

\begin{figure}
\begin{center}
\includegraphics[width=8cm]{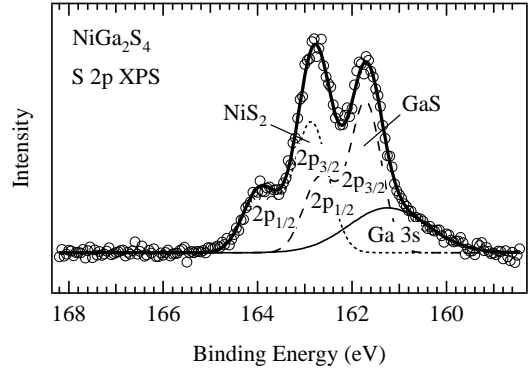}
\caption{S 2$p$ and Ga 3$s$ core-level XPS of NiGa$_2$S$_4$
(open circles).
The spectrum is fitted to five Gaussians: the Ga 3$s$ component
(thin solid curve), the 2$p_{3/2}$ and 2$p_{1/2}$
components of the GaS layer (dashed curve), 
and those of the NiS$_2$ layer (dotted curve).
The fitted result is shown by the thick solid curve.}
\label{S2p}
\end{center}
\end{figure}

The S 2$p$ core-level photoemission spectrum of NiGa$_2$S$_4$
is shown in Fig. \ref{S2p}. The S 2$p$ spectrum can be	
decomposed into four components: the 2$p_{3/2}$ and 2$p_{1/2}$
components of the GaS layer and those of
the NiS$_2$ layer. The S 2$p$ peaks of the NiS$_2$ layer
are higher in binding energy than those of the GaS layer.
The energy difference of $\sim$ 1 eV indicates that
the S ions in the NiS$_2$ layer have less valence
(S 3$p$) electrons than those in the GaS layer.
This observation already suggests that the
amount of the S 3$p$ holes in the NiS$_2$ layer
is substantial as revealed by the Ni 2$p$ spectrum
in the next paragraph.

\begin{figure}
\begin{center} 
\includegraphics[width=8cm]{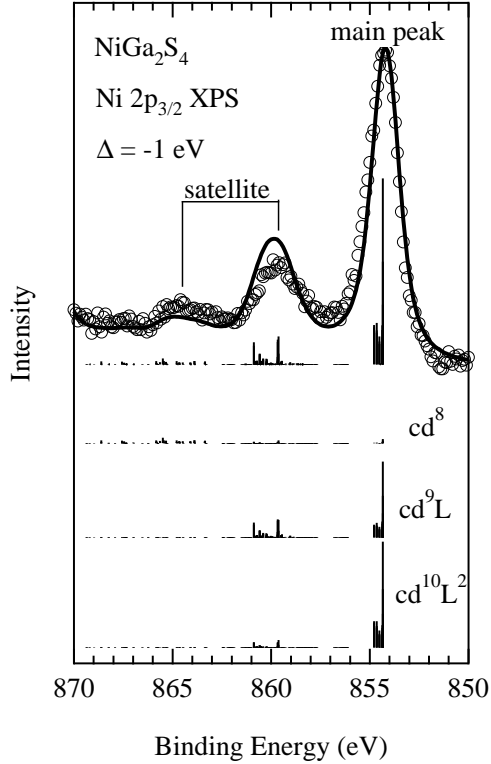}
\caption{Ni 2$p$ core-level XPS of NiGa$_2$S$_4$
(open circles).
The calculated line spectrum is broadened 
(solid curve) and is compared with the experimental
result. In the lower panel,
the line spectrum is decomposed into the
$cd^8$, $cd^{9}L$, and $cd^{10}L^2$ components.}
\label{Ni2p}
\end{center}
\end{figure}

The Ni 2$p$ core-level spectrum of NiGa$_2$S$_4$
is shown in Fig. \ref{Ni2p}. The Ni 2$p_{3/2}$ spectrum
consists of three structures: the main peak at $\sim$ 854 eV
and the satellite structures at $\sim$ 859 eV and 864 eV.
In order to extract the electronic structure parameters
such as the S 3$p$ to Ni 3$d$ charge-transfer energy $\Delta$,
the Coulomb interaction between the Ni 3$d$ electrons $U$,
and the transfer integrals between the S 3$p$ and Ni 3$d$
orbitals $(pd\sigma)$, we have performed configuration-interaction
calculations using an octahedral NiS$_6$ cluster model
(see Fig. \ref{structure}).
Since the NiS$_6$ octahedra form the NiS$_2$ layer 
sharing their edges, the interaction between 
the neighboring NiS$_6$ octahedra via the S 3$p_{\sigma}$
orbitals pointing to the Ni sites is rather weak compared
to that in the corner sharing case \cite{Veenendaal}. 
Therefore, the cluster-model analysis is expected to give
a good description of the Ni 2$p$ core-level spectrum.
In the present cluster model, the Coulomb interaction 
between the Ni 3$d$ electrons are given by the Slater
integrals $F^0(3d,3d)$, $F^2(3d,3d)$, and $F^4(3d,3d)$.
The average Ni 3$d$-Ni 3$d$ Coulomb interaction $U$
is expressed by $F^0(3d,3d)$ and is an adjustable parameter.
$F^2(3d,3d)$ and $F^4(3d,3d)$ are fixed to 80\% of 
the atomic Hartree-Fock values \cite{deGroot}.
The Coulomb interaction
between the Ni 2$p$ core hole and the Ni 3$d$ electron
is expressed by the Slater integrals $F^0(2p,3d)$, 
$F^2(2p,3d)$, and $G^1(2p,3d)$. 
The average Ni 2$p$-Ni 3$d$ Coulomb interaction $Q$
is expressed by $F^0(2p,3d)$ and is fixed to $U/0.8$ 
\cite{Bocquet, Okada}.
$F^2(2p,3d)$ and $G^1(2p,3d)$ are fixed to 80\% of 
the atomic Hartree-Fock values \cite{deGroot}.

In Fig. \ref{Ni2p}, the calculated line spectrum is 
broadened and compared with the experimental result.
The three-peak structure of 
the Ni 2$p_{3/2}$ spectrum is well reproduced
by the calculation using $\Delta$=-1.0 eV, $U$=5.0 eV,
and $(pd\sigma)$=-1.0 eV.
The ground state is given by the linear 
combination of $d^8$, $d^9L$, and $d^{10}L^2$
configurations:
\begin{equation}
\Psi_g = \alpha|d^8\rangle + \beta|d^{9}L\rangle
+ \gamma|d^{10}L^2\rangle,
\end{equation}
where $L$ denotes a hole in the S 3$p$ orbitals.
The final states are given by the linear 
combinations of $cd^8$, $cd^9L$, and $cd^{10}L^2$
configurations, where $c$ denotes a Ni $2p$ core hole.
In the final states,
the Coulomb interaction between the Ni $2p$ core hole 
and Ni $3d$ electrons stabilizes $cd^{10}L^2$ compared with $cd^9L$.
The present analysis gives $\alpha^2=0.25$,
$\beta^2=0.60$, and $\gamma^2=0.15$,
and the ground state is dominated by
the $d^9L$ configuration.
Negative charge-transfer energy compounds
have been found in some transition-metal oxides 
with high valence such as Cu$^{3+}$ and Fe$^{4+}$
\cite{Mizokawa}.
The present system is the first example
of transition-metal sulfide with negative
charge-transfer energy.
The negative charge-transfer energy indicates that 
the ground state has strong S 3$p$ hole character.
Although the symmetry of the ground state is $^3A_{2g}$,
the magnitude of the local spin at the Ni site is
reduced to 0.55 in comparison with the ionic value $S=1$ 
by the strong Ni 3$d$-S 3$p$ hybridization.
In this sense, the $^3A_{2g}$ state with the $d^9L$ character
is a triplet analogue of the Zhang-Rice singlet
in high-Tc cuprates.

\begin{figure}
\begin{center} 
\includegraphics[width=8cm]{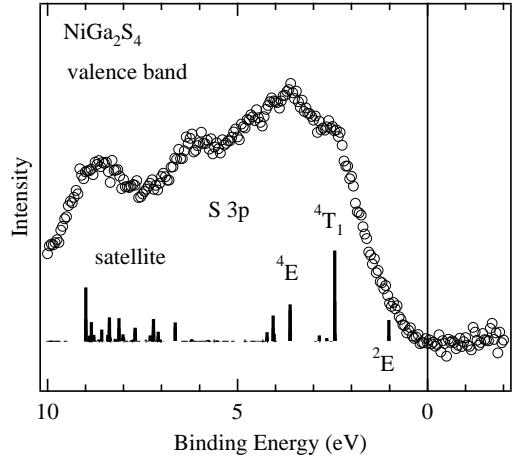}
\caption{Valence-band XPS of NiGa$_2$S$_4$ (open circles).
The line spectrum is obtained by the NiS$_6$
cluster model calculation.}
\label{val}
\end{center}
\end{figure}

In Fig. \ref{val}, the valence-band spectrum is plotted
with the line spectrum obtained by the NiS$_6$ cluster
model calculation with the parameters obtained from
the Ni 2$p$ spectrum. Although the contribution from
the GaS layer prevents us from fitting the valence-band
spectrum, the overall agreement between the line spectrum 
and the valence-band spectrum support the cluster-model
analysis of the Ni $2p$ spectrum.
The first ionization state has the symmetry of $^2E_g$
and the second ionization state is $^4T_{1g}$.
This situation is similar to those of NiS and NiS$_2$
which have small but positive charge-transfer energy
\cite{NiS, NiS2}. 
The excitation from the $^3A_{2g}$ ground state to
the $^2E_g$ first ionization state is obtained by 
removal of an $e_g$ electron. The $^2E_g$ state
is expected to have some energy-momentum dispersion 
due to the interaction between the NiS$_6$ clusters
and to form the $e_g$ band. 

\begin{figure}
\begin{center} 
\includegraphics[width=6cm]{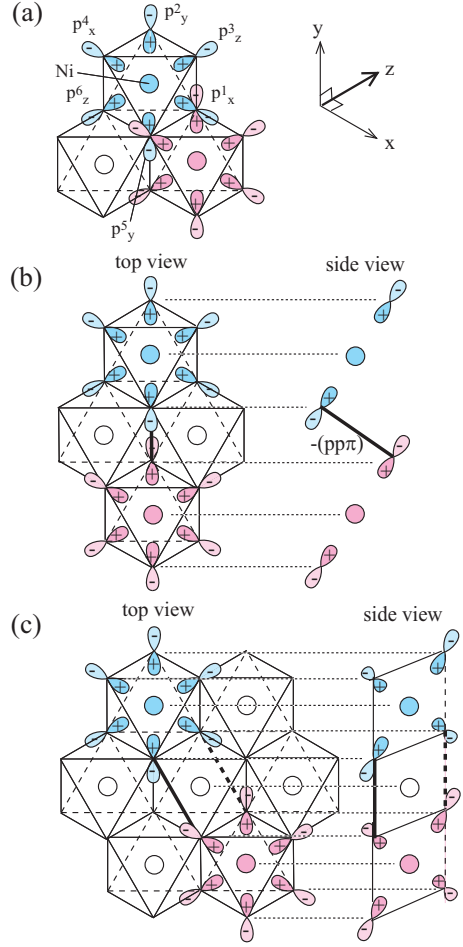}
\caption{(Color online) (a) Top view of the nearest neighbor
clusters in the NiS$_2$ layer.
(b) Top and side views for superexchange pathway 
bewteen the second nearest neighbor clusters, which is given
by the S 3$p$-3$p$ transfer of $-(pp\pi)$, as indicated by the thick solid line.
(c) Top and side views for two superexchange pathways bewteen the third nearest
neighbor clusters, which are given by the S 3$p$-3$p$ transfer 
of $[-(pp\sigma)+(pp\pi)]/\sqrt{2}$, as indicated by the thick solid
and broken lines.
}
\label{superexchange}
\end{center}
\end{figure}

In the next step, let us examine superexchange pathways
in NiGa$_2$S$_4$ based on the electronic structure parameters
obtained from the analysis of the Ni $2p$ spectrum. 
Since the $t_{2g}$ orbitals are fully occupied and 
the neighboring Ni-S bonds are approximately orthogonal 
to each other in the two NiS$_6$ clusters sharing their edges, 
the cell perturbation calculation \cite{Jefferson}
becomes rather simple and 
the superexchange interaction between the two clusters is 
given by the second order perturbation of S 3$p$-3$p$ transfer term.
In the NiS$_6$ cluster, the exact ground state
is given by $\Psi_g(^3A_{2g}) = \alpha|d^8(^3A_{2g})\rangle
 + \beta|d^{9}L(^3A_{2g})\rangle + \gamma|d^{10}L^2(^3A_{2g})\rangle$. 
Here, the NiS$_6$ cluster is assumed to be a cubic octahedron in the calculation
although it has a slight trigonal distortion in NiGa$_2$S$_4$.
Since the S 3$p$ hole denoted as $L$ is dominated by 
the 3$p_{\sigma}$ orbitals pointing to the Ni sites,
it is reasonable to approximate that
\begin{equation}
|d^{9}L(^3A_{2g})\rangle=\frac{1}{\sqrt{2}}
(|d_{x^2-y^2}L_{3z^2-r^2}|+|L_{x^2-y^2}d_{3z^2-r^2}|).
\end{equation}
Here, $d_{x^2-y^2}$ and $d_{3z^2-r^2}$ denote a hole
in Ni 3$d$ $x^2-y^2$ and $3z^2-r^2$ orbitals, respectively.
The $x$-, $y$-, and $z$-axes are along the Ni-S bonds 
as shown in Fig. \ref{superexchange}(a).
The $x^2-y^2$ type S 3$p$ molecular orbital is given by
$L_{x^2-y^2}=\frac{1}{2}(p^1_x-p^2_y+p^4_x-p^5_y)$,
in which $p^1_x$, $p^2_y$, $p^4_x$, and $p^5_y$
are $p_x$ and $p_y$ orbitals pointing to the Ni site
[see Fig. \ref{superexchange}(a)].
The $d_{3z^2-r^2}$ type S 3$p$ molecular orbital is given by
$L_{3z^2-r^2}=\frac{1}{\sqrt{3}}(p^3_z+p^6_z)
-\frac{1}{2\sqrt{3}}(p^1_x+p^2_y+p^4_x+p^5_y)$,
in which $p^3_z$ and $p^6_z$ are $p_z$ orbitals 
pointing to the Ni site. As a consequence,
the overlap integral between the neighboring $d^9L$
states with $^3A_{2g}$ symmetry can be neglected. 
The superexchange pathways 
between the neighboring $d^9L(^3A_{2g})$ states
are given by the transfer terms between 
the two S 3$p$ molecular orbitals with $L_{x^2-y^2}$
and $L_{3z^2-r^2}$.
The transfer integrals
between the neighboring $L_{x^2-y^2}$ orbitals
is $\frac{1}{4}[(pp\sigma)+(pp\pi)]$ and
that between the neighboring $L_{3z^2-r^2}$ orbitals
is $\frac{1}{12}[(pp\sigma)+9(pp\pi)]$, where 
$(pp\sigma)$ and $(pp\pi)$ are the S 3$p$-3$p$ transfer integrals.
Therefore, the antiferromagntic superexchange interaction 
between the neighboring sites is given by 
\begin{equation}
J_1^{AF} = -\frac{\beta^2}{2U_p}
\{\frac{1}{16}[(pp\sigma)+(pp\pi)]^2
+\frac{1}{144}[(pp\sigma)+9(pp\pi)]^2\},
\end{equation}
where $U_p$ is the Coulomb interaction between 
the S 3$p$ holes at the same S sites and is
expected to be $\sim$ 1 eV \cite{Up}. 
Since  $\beta^2 = 0.6$, $(pp\sigma)$ = 0.6 eV, 
and $(pp\pi)$ = -0.15 eV, $J_1^{AF}$ $\sim$ -5 meV.
On the other hand, the two S 3$p$ holes
at the shared sulfur sites have Hund coupling
due to the on-site exchange integral $J_p$
at the S site. The ferromagnetic interaction
due to the Hund coupling is given by 
$J_1^{F} = \frac{\beta^4}{18}J_p$ $\sim$ 4 meV,
where $J_p$ is the Hund coupling between 
the S 3$p$ holes at the same S sites and 
is assumed to be $\sim$ 0.2 eV.
As a result, the total magnetic interaction 
between the neighboring sites $J_1=J_1^{AF}+J_1^{F}$  
is estimate to be $\sim$ -1 meV.
The transfer integral between 
the second neighbor $L_{x^2-y^2}$ orbitals 
is given by $-\frac{1}{3}(pp\pi)$ and that
between the second neighbor $L_{3z^2-r^2}$ orbitals
can be neglected [see Fig. \ref{superexchange} (b)]. 
The superexchange interaction $J_2$ due to the transfer 
term between the second neighbor $d^9L$ states becomes
\begin{equation}
J_2=-\frac{\beta^2}{18U_p}(pp\pi)^2,
\end{equation}
which is $\sim$ -1 meV and as small as $J_1$.

In contrast,
the transfer integral between the third neighbor 
$L_{x^2-y^2}$ orbitals is $\frac{1}{4}[(pp\sigma)-(pp\pi)]$ and
that between the third neighboring $L_{3z^2-r^2}$ orbitals
is $\frac{1}{4\sqrt{3}}[-(pp\sigma)+(pp\pi)]$
as shown in Fig. \ref{superexchange} (c). 
The superexchange interaction $J_3$ between
the third neighbor sites due to the transfer 
term between the $d^9L$ states is given by 
\begin{equation}
J_3=-\frac{\beta^2}{24U_p}[(pp\sigma)-(pp\pi)]^2,
\end{equation}
which is estimated to be $\sim$ -14 meV.
The superexchange interaction between the forth neighbor
sites is beyond the second order perturbation on the
$p$-$p$ transfer integrals $(pp\sigma)$ and $(pp\pi)$,
and is expected to be negligibly small.
The present cell perturbation analysis predicts that
the third neighbor superexchange interaction is
dominant consistent with the neutron diffraction result
\cite{Nakatsuji}.

In conclusion, the electronic structure of NiGa$_2$S$_4$ 
with the NiS$_2$ triangular lattice is investigated 
using photoemisssion experiment and model calculations.
The Ni 2$p$ photoemission data and the cluster model 
calculation show that the ground state has the $d^9L$ character
($L$ is a S 3$p$ hole) which is a triplet analogue of
the Zhang-Rice singlet state.
The strong S 3$p$ hole character of the ground state
provides the enhanced superexchange interaction between 
the third nearest neighbor sites. NiGa$_2$S$_4$ is
a unique spin-disordered system in that the negative
charge-transfer energy allows unexpectedly long 
superexchange pathways. 

This work was supported by Grant-In-Aid from
Ministry of Education, Culture, Sports, 
Science and Technology of Japan.

\end{document}